\documentclass[lettersize,journal]{IEEEtran}
\usepackage{amsmath,amsfonts}
\usepackage{algorithmic}
\usepackage{algorithm}
\usepackage{array}
\usepackage[caption=false,font=normalsize,labelfont=sf,textfont=sf]{subfig}
\usepackage{textcomp}
\usepackage{stfloats}
\usepackage{url}
\usepackage{verbatim}
\usepackage{graphicx}
\usepackage{cite}
\usepackage{booktabs}
\usepackage{caption}
\usepackage{tikz}
\usepackage{multirow}
\usepackage{siunitx}
\begin{document}

\title{Practical Semantic Communication for General-Purpose Networks: Compression of High-Dimensional Embeddings Under Quantization}

\author{

\IEEEauthorblockN{Leonardo Roque{\textsuperscript{*,$\dagger$}}, Louis-Adrien Dufrène{\textsuperscript{*}}, Quentin Lampin{\textsuperscript{*}}, Grégoire Lefebvre{\textsuperscript{*}}, Mohamad Assaad{\textsuperscript{$\dagger$}}}\\
    \IEEEauthorblockA{*Orange Research, France\\
    $\dagger$Laboratoire des Signaux et Systèmes, CentraleSupélec, France\\
    Email: \{leonardo.roquealmeidamatos, louisadrien.dufrene, quentin.lampin,  gregoire.lefebvre\}@orange.com, \\mohamad.assaad@centralesupelec.fr}

\thanks{This research was funded in part by the Agence Nationale de la Recherche (ANR) under the ANR-24-IAS1-0002-02 COMSEMA project and the SNS
JU 6GARROW project under the EU’s Horizon program Grant Agreement
No 101192194, in collaboration with ROK.}%
}

\maketitle

\begin{abstract}
This paper introduces a practical framework for semantic communication over wireless channels, focusing on the compression of high-dimensional embeddings under quantization and channel errors. Building upon a relaxed Gaussian Information Bottleneck (GIB) model, the approach provides analytical solutions for task-relevant information preservation, even with non-Gaussian targets. The framework relaxes traditional Gaussian assumptions, enabling effective compression for discrete and complex data such as text embeddings. It derives closed-form gradients for optimizing the compression matrix and establishes bounds linking continuous representations to discrete, noisy system performance. Comprehensive experiments on text classification tasks demonstrate that the proposed method outperforms task-agnostic quantization-based compression methods, maintaining high semantic fidelity at low bit budgets per embedding. The approach offers a scalable, interpretable, and resource-efficient solution for deploying semantic communication in general-purpose networks.
\end{abstract}

\begin{IEEEkeywords}
Semantic Communication, Wireless Network, Gaussian Information Bottleneck, Massive Text Embedding Benchmark.
\end{IEEEkeywords}

\section{Introduction}
\IEEEPARstart{T}{he} emergence of semantic communication represents a paradigm shift in wireless system design, moving beyond the traditional communication framework of reliable bit transmission toward task-oriented information delivery~\cite{shannon_mathematical_1948,qin2022semanticcommunicationsprincipleschallenges,gunduz_beyond_2023}. In this new paradigm, the objective is focused on preserving the information necessary for completing a specific task, such as binary or multiclass text classification. This fundamental change enables dramatic improvements in spectral efficiency, particularly in bandwidth-constrained wireless networks~\cite{xie_deep_2021}.

To support large-scale connectivity and efficient task-oriented transmission, semantic communication has been investigated extensively in recent years. Deep joint source-channel coding (DJSCC) approaches have demonstrated promising results by training neural encoder-decoder pairs end-to-end to directly map source data to channel inputs. Bourtsoulatze et al.~\cite{bourtsoulatze_deep_2019} pioneered deep JSCC for wireless image transmission, achieving superior rate-distortion performance compared to separate source-channel coding. Recent work has further extended this framework through transformer-based architectures, such as WITT~\cite{yang2023witt} for wireless image semantic transmission, as well as generative semantic communication approaches that leverage powerful generative models for semantic reconstruction and content delivery~\cite{Liang2023GenerativeAS}.

To extend the scope beyond single-task systems, multi-task and task-agnostic semantic communications have been explored, with the 
overarching goal of designing a single, unified framework 
that generalizes across heterogeneous downstream tasks without 
retraining. Early work by Xie et al.~\cite{xie2022task} proposed multiuser semantic communication systems for visual question answering by jointly processing text and image data. More recently, contrastive learning-based semantic communication frameworks have been developed to learn semantically meaningful representations that remain robust under channel impairments, while reducing the dependence on explicitly task-coupled supervision~\cite{tang2024contrastive}. Zhang et al.~\cite{zhang2024unified} further investigated unified multi-task frameworks for multimodal data transmission, addressing the challenge of handling heterogeneous data modalities within a single pipeline.

Foundation model-based approaches have emerged as a major trend in 
semantic communication research, motivated by the observation that 
large pre-trained models already encode rich, generalizable semantic 
representations that can be directly leveraged for communication 
purposes. Rather than training task-specific encoders from scratch, 
these approaches exploit the broad knowledge embedded in foundation 
models to build semantic encoders that are both versatile and 
data-efficient, reducing the training overhead compared 
to conventional deep learning-based semantic communication systems. Recent work has investigated large model-empowered 
semantic communication architectures and the direct use of large 
language models as semantic communication engines, highlighting their 
potential to support flexible deployment and reduced retraining 
requirements across diverse tasks~\cite{large_model_semcom2024,llm_enabled_semcom2024}.

Information-theoretic approaches have provided theoretical foundations 
for semantic communication design, addressing a fundamental limitation 
of purely data-driven systems: the absence of formal guarantees on 
compression quality and task performance. Concretely, the information bottleneck (IB) principle~\cite{tishby2000information} 
offers a theoretically grounded framework for extracting minimal yet 
sufficient semantic representations by systematically discarding 
task-irrelevant information while preserving features critical to 
downstream tasks. A notable analytical instantiation of this principle 
is the Gaussian Information Bottleneck (GIB)~\cite{chechik2005information}, 
which yields optimal closed-form solutions under jointly Gaussian assumptions, 
offering interpretable parameters and minimal computational cost.
In parallel, rate-distortion inspired formulations have also been explored 
for semantic communication, particularly in multi-task settings where the 
trade-off between compression efficiency and task utility must be 
characterized explicitly~\cite{multitask_rd_semcom2023}.

Despite these advances, deploying semantic communication systems in practical, general-purpose networks introduces a distinct and largely unaddressed set of 
operational constraints. These systems must support 
dynamic and unpredictable traffic patterns~\cite{hüttebräucker2024pragmatic}, accommodate a wide 
diversity of downstream tasks without prior knowledge of which tasks 
will be requested~\cite{lampin2024semantic,matos2025compacting}, and operate under strict quality-of-service requirements. 
Beyond device heterogeneity, operational networks require modularity: new tasks or services must be deployable without 
redesigning the entire communication pipeline, in which case the dynamic adaptation of network functionalities would be impractical in real-world scenarios. Finally, the transition 
from continuous learned representations to digital communication introduces quantization 
constraints that must be accounted for at the system design 
level. These constraints collectively define a demanding operational envelope, posing challenges that existing semantic communication frameworks have yet to address.

Existing approaches address some of these challenges but 
fall short in key aspects when evaluated against this full operational 
envelope. DJSCC and foundation model-based methods achieve impressive 
performance but remain either highly task-specific or computationally 
prohibitive for resource-constrained 
devices~\cite{bourtsoulatze_deep_2019,yang2023witt,Liang2023GenerativeAS,
large_model_semcom2024,llm_enabled_semcom2024}. Multi-task approaches 
partially address task diversity but require joint training across all 
target tasks simultaneously, limiting scalability when new tasks emerge 
dynamically~\cite{xie2022task,tang2024contrastive,zhang2024unified}. From 
a theoretical standpoint, information-theoretic approaches offer the 
performance guarantees that operational deployment requires, yet each 
existing framework faces fundamental limitations in this context. 
Rate-distortion-based formulations provide useful insight into 
compression--utility trade-offs but often remain too abstract to 
translate directly into finite-dimensional practical pipelines with 
quantization and channel 
perturbations~\cite{multitask_rd_semcom2023}. The classical Gaussian 
Information Bottleneck~\cite{chechik2005information} provides 
closed-form analytical solutions with minimal computational cost, but 
its Gaussian assumptions fundamentally limit direct applicability to 
tasks with discrete labels, which are central in practical semantic 
communication.

Among the existing frameworks, Variational Information Bottleneck (VIB) and Vector Quantization-Variational Autoencoder (VQ-VAE) based approaches stand out as the most compatible with the aforementioned operational constraints. VIB methods~\cite{alemi2017deep} naturally accommodate more plural data parameterizations for general-purpose networks from neural semantic encoding/decoding and have shown usefulness under distribution shifts~\cite{li2024distribution_shift_ib}, but do not explicitly account for quantization and still require a data-dependent neural training architecture to ensure encoder expressivity and accurate variational approximations.
VQ-VAE-based approaches produce discrete representations 
that are directly compatible with digital communication interfaces~\cite{oord2017neural,
hu2023robustsemanticcommunicationsmasked}, but their training objective is fundamentally reconstruction-based 
— optimizing input fidelity rather than task-relevant information 
preservation — and therefore provides no analytical insight 
on downstream task performance under quantization and channel 
errors. Taken together, no existing framework jointly provides analytical tractability to preserve task quality, compatibility with discrete targets, and explicit accounting for the operational constraints of practical networks.

To bridge this identified gap in information-theoretic approaches, this paper develops a practical, deployable framework for semantic communication over wireless channels that addresses operational challenges while maintaining theoretically grounded efficiency. Based on a relaxed Gaussian Information Bottleneck configuration, our proposal combines analytical tractability with distortion guarantees to improve compression, and general-purpose network operability. Our goal is not theoretical optimality in the pure information-theoretic sense, but rather a realistic, implementable compression framework that adapts to operational network constraints, performs a specialized task-oriented feature extraction to maintain good task performance, and enables flexible deployment.

The main contributions of this paper are:
\begin{enumerate}
    \item Relaxed Gaussian assumptions: We extend the 
    original GIB formulation to require only class-conditional 
    Gaussianity of the source given the target. The marginal source 
    distribution is unrestricted, and the target is discrete, making 
    the method directly applicable to classification. The source marginal 
    is then a Gaussian mixture, in contrast to the single joint Gaussian 
    required by classical GIB.
    \item Distortion-based theoretical foundation: We 
    establish a rigorous bound relating the continuous GIB 
    optimization objective to the practical discrete objective 
    under quantization, providing the theoretical justification 
    for connecting the continuous compressed variable to its 
    quantized representation at the receiver.
    \item Analytical gradient derivation: We derive 
    closed-form expressions for the gradients of the discrete GIB 
    objective under relaxed assumptions, enabling efficient 
    compression optimization.
    \item Practical optimization framework: We develop a 
    complete algorithm integrating continuous optimization with 
    K-Means quantization design and iterative codebook updates to 
    minimize distortion while preserving task-relevant information.
    \item Experimental validation: Comprehensive experiments 
    on the Banking77 dataset demonstrate consistent outperformance 
    over task-agnostic baselines across all evaluated compression 
    configurations.
\end{enumerate}

\section{Background and Problem Formulation}
\label{sec:background}

\subsection{Information Bottleneck Principle}
The Information Bottleneck (IB) principle, introduced by Tishby, Pereira, and Bialek~\cite{tishby2000information}, provides an information-theoretic framework for extracting relevant information from data while discarding irrelevant details. At its core, the IB method seeks to find a compressed representation $T$ of an input variable $X$ that preserves maximal information about a target variable $Y$. This is formalized through the optimization of a trade-off between compression and relevance:
\begin{equation}
    \min_{p(t|x)} \left[ I(X;T) - \beta I(T;Y) \right],
\end{equation}

where $I(X;T)$ measures the compression cost, $I(T;Y)$ quantifies the preserved relevant information, and $\beta$ controls the trade-off between these competing objectives. This framework has proven fundamental for understanding representation learning~\cite{bengio_representation_2013,shamir2010learning}, providing a principled approach to feature extraction, dimensionality reduction, and task-oriented compression.

\subsection{Gaussian Information Bottleneck}
\label{subsec:GIB}
The Gaussian Information Bottleneck (GIB), developed by Chechik et al.~\cite{chechik2005information}, specializes this framework to the case where the source $X$ and the target $Y$ are jointly Gaussian, enabling analytical solutions and providing deep insights into optimal compression under second-order statistics. In the GIB formulation, the compressed representation is obtained through a linear transformation with additive Gaussian noise:
\begin{equation}
    T = AX + Z, \quad Z \sim \mathcal{N}(0, \Sigma_Z),
\end{equation}

where $A$ is the compression matrix to be optimized, and $Z$ represents controlled noise injection that limits the information preserved about X. This continuous formulation has proven valuable for understanding fundamental limits of lossy compression~\cite{cover1999elements}.

\subsection{Related Approaches and Limitations}
Several approaches have been proposed to address limitations of classical GIB. VIB~\cite{alemi2017deep} relaxes distributional assumptions by parameterizing the encoder as a neural network, enabling application to arbitrary data distributions and discrete targets. However, VIB demands end-to-end training of neural encoder architectures for optimizing variational bounds on mutual information, necessitating substantial data to ensure encoder expressivity and good distribution approximations. Recent extensions include nonlinear IB methods~\cite{kolchinsky2019nonlinear} that assume arbitrarily distributed $X$ and $Y$ while allowing nonlinear transformations, but still relying on the optimization of a non-parametric upper bound for mutual information and neural network-based encoding; and discrete IB approaches~\cite{strouse2017deterministic} that directly optimize discrete representations but require combinatorial optimization.

Quantization-aware methods have been explored to bridge continuous 
semantic representations and the discrete inputs required by digital communication channels. The authors of~\cite{vq_semcom2023} proposed a vector-quantized semantic communication framework to map learned semantic features to discrete codebook representations, while the work in~\cite{matos2025compacting} investigated the compression efficiency of product quantization for high-dimensional embeddings. In parallel, vector quantization-based 
approaches inspired by the VQ-VAE framework~\cite{oord2017neural} have emerged as a promising alternative, offering structured discrete latent representations that are jointly learned with the semantic encoder. The authors of~\cite{hu2023robustsemanticcommunicationsmasked} proposed a masked VQ-VAE with adversarial training to build robust semantic codebooks resilient to semantic noise, while the work in~\cite{lyu2025vqvaebaseddigitalsemantic} demonstrated the compatibility of VQ-VAE-based discrete representations with practical digital communication systems through an importance-aware OFDM transmission strategy. However, these approaches generally require training on large datasets and do not provide formal theoretical analysis on the relationship between quantization distortion and task performance degradation, which is a fundamental perspective in the design of a semantic communication system.

In this work, we extend the classical GIB to accommodate practical semantic communication while providing a theoretically grounded strategy to perform dedicated task-oriented feature extraction. Particularly, we are interested in assessing performances of the proposed framework against the task-agnostic compression proposed in \cite{matos2025compacting}, which is not designed to retain the most relevant features to the downstream task.

\subsection{Practical Implementation Challenges}
\label{impl_challenges}
Despite its theoretical elegance and analytical tractability, a fundamental gap exists between the continuous GIB framework and practical digital communication systems. Real-world implementations face three critical challenges that the standard GIB formulation does not address.

First, an important limitation of the standard GIB formulation is the requirement that $(X, Y)$ be jointly Gaussian~\cite{chechik2005information}. This assumption, while enabling analytical tractability, is highly restrictive in practice. Joint Gaussianity implies linear relationships between $X$ and $Y$ and Gaussian marginal distributions for both variables, excluding many practical scenarios, most notably classification tasks where the target $Y$ represents discrete class labels.

Second, quantization is unavoidable in general-purpose wireless networks. Practical systems cannot transmit, store, or process continuous-valued representations. Any real implementation must discretize the compressed representation $T$ into a finite set of symbols, which are then encoded into binary sequences for transmission or storage. This quantization process fundamentally changes the information-theoretic properties of the system, as the relevant quantity becomes the mutual information between the discrete quantized representation and the target, not the continuous representation.

Third, transmission occurs over noisy channels. In communication systems, the binary-encoded representation must traverse imperfect channels subject to noise, and errors. These channel impairments corrupt the transmitted bits, and the receiver must work with a degraded, noisy version of the intended representation.

The practical data flow in a communication system becomes:
\begin{equation}
    X \rightarrow T \rightarrow B \rightarrow \hat{B} \rightarrow Y,
\end{equation}

where:

$X$ is the source variable;
$T = AX + Z$ is the continuous compressed representation;
$B \in \{0,1\}^\ell$ is the binary index of the quantization centroids obtained by partitioning the space of $T$, where $\ell \geq \lceil \log_2 L\rceil$ and $L$ represents the number of quantization centroids;
$\hat{B}$ is the noisy received bit sequence (e.g., after transmission through a Binary Symmetric Channel with a crossover probability $p_{error}$);
and $Y$ is the target variable to be inferred from $\hat{B}$.

\subsection{Problem Formulation}
The true objective of a practical communication system should be:
\begin{equation}
    \mathcal{L}(A) = I(X;\hat{B}) - \beta I(\hat{B};Y),
    \label{eq:practical-IB-objective}
\end{equation}

where $\hat{B}$ is the actual noisy received representation available at the receiver. This objective captures both compression efficiency (how much information about $X$ survives quantization and channel noise) and task relevance (how much information about $Y$ is preserved in the noisy received bits).

The fundamental challenge is that while the objective function \eqref{eq:practical-IB-objective} involves discrete, noisy random variables ($\hat{B}$), the optimization variable is the continuous matrix A. This creates several profound difficulties.

First, the discrete mutual information terms are intractable. Computing $I(X;\hat{B})$ requires:
\begin{equation}
    I(X;\hat{B}) = \sum_{\hat{b}} \int p(x)p(\hat{b}|x) \log \frac{p(\hat{b}|x)}{p(\hat{b})} dx,
\end{equation}

where:
\begin{equation}
    p(\hat{b}|x) = \sum_b p(\hat{b}|b) \int_{\mathcal{C}_b} p(t|x) dt.
\end{equation}

Here, $\mathcal{C}_b$ is the quantization region mapping to binary index $b$, and $p(\hat{b}|b)$ is the channel transition probability. This involves integration over complex, high-dimensional quantization regions, summation over all $2^\ell$ possible transmitted and received bit patterns, and accounting for stochastic channel transitions.

Second, gradient computation is highly non-trivial. To optimize $A$ using gradient descent, we need:
\begin{equation}
    \nabla_A \mathcal{L}(A) = \nabla_A I(X;\hat{B}) - \beta \nabla_A I(\hat{B};Y),
    \label{eq:IB-gradient}
\end{equation}

The dependence of $I(X;\hat{B})$ on $A$ is indirect and complex, passing through the continuous distribution $p(t|x)$, the quantization regions $\mathcal{C}_b$, the integration over these regions, the channel transition probabilities, and the summation over bit patterns.

\subsection{Main Contributions}
To address the discussed limitations, this paper presents a compression framework that extends the GIB to practical semantic communication systems, accounting for quantization and channel noise while providing closed-form expressions to optimize compression without requiring end-to-end training or strong Gaussian assumptions.

First, we establish a distortion-based theoretical foundation that rigorously connects the continuous optimization objective to practical system performance. This bound is essential because our gradients optimize the properties of $T$ (through $A$), but the actual system performance depends on $\hat{B}$ (the reconstruction from noisy bits). The bound guarantees that if distortion is kept small through careful quantization design, then improving $I(T;Y)$ through optimization of $A$ will improve the performance $I(\hat{B};Y)$.

Second, we derive closed-form analytical expressions for the gradients of the discrete, noisy IB objective \eqref{eq:IB-gradient}. The key technical contribution is showing that these gradients can be computed by combining analytical gradients of the continuous Gaussian distributions with numerical integration over quantization regions and summation over channel transition probabilities. In this part, we also demonstrate that we can relax the joint Gaussianity requirement dramatically. We show that our analytical gradient derivations remain valid under the minimal assumption of class-conditional Gaussianity, $X|Y=y \sim \mathcal{N}(\mu_{X|y}, \Sigma_{X|y})$, for a discrete target $Y$. Critically, we do not require:

\begin{enumerate}
    \item $Y$ to be Gaussian,
    \item $(X, Y)$ to be jointly Gaussian,
    \item $X$ to be Gaussian: its marginal $p(x)=\sum_y p(y)p(x|y)$ is generally a Gaussian mixture, and the distribution $p(y)$ is arbitrary.
\end{enumerate}

This relaxation is possible because our gradient derivations depend only on the conditional distributions $p(T|X) = \mathcal{N}(AX, \Sigma_Z)$ and $p(T|Y=y) = \mathcal{N}(A\mu_{X|y}, S)$, where $S = A\Sigma_{X|y}A^T + \Sigma_Z$ depends on the class $y$. These conditional distributions remain Gaussian under our assumptions, and no Gaussian model of the marginal $p(x)$ appears in the gradient expressions.

This has profound practical implications. For text classification, features extracted from neural networks are often modeled as approximately Gaussian within each class, even though the relationship between features and class labels is highly non-linear. Our framework applies directly to such scenarios without requiring any transformation of discrete labels.

Finally, by integrating continuous optimization, quantization design, and distortion minimization, the designed algorithm operates in two stages: 1) during a first training step, optimize the construction of the compression matrix $A$ using the analytical gradients to minimize \eqref{eq:practical-IB-objective}, and 2) apply the optimized compression matrix on the original embeddings and assess task performances from using these compressed representations after quantization and transmission through a noisy channel. By the derived distortion bound, we ensure that the practical system performance $I(\hat{B};Y)$ approaches the optimized $I(T;Y)$.

\section{System Model}
\label{sec:system}
A key enabler of semantic communication is the use of learned embeddings—high-dimensional, real-valued vector representations that capture the semantic content of data in a structured, continuous space~\cite{mikolov_efficient_2013,devlin_bert_2019}. Modern machine learning systems naturally operate with such embeddings as multimodal data representations, and these have emerged as a de facto standard for semantic data representation across diverse applications~\cite{he_deep_2016,dosovitskiy_image_2021,covington_deep_2016}. The embeddings' generality of representation on multimodal data make them attractive for semantic communication systems, as a single communication protocol designed for embedding transmission can support multiple downstream tasks without necessitating specific network adaptations. In this perspective, these structured vectors can serve then as a standard, universal data format for modern communications~\cite{lampin2024semantic}, allowing to encode a variety of data types while preserving their intrinsic semantic characteristics.

However, embeddings are inherently high-dimensional—typical dimensions range from hundreds to thousands—creating a fundamental challenge in wireless semantic communication. Transmitting these high-dimensional representations over bandwidth-limited wireless channels can consume substantial network resources, particularly when accounting for the overhead of reliable transmission over noisy channels~\cite{bengio_representation_2013,bourtsoulatze_deep_2019}. This motivates the compression-oriented system design described below.

\subsection{Semantic Communication System Architecture}
\begin{figure*}[t!]
\centering
\resizebox{\textwidth}{!}{%
\begin{tikzpicture}[thick,
    block/.style={draw, rectangle, rounded corners, 
                  minimum height=2.2cm, minimum width=3.8cm,
                  align=center, font=\LARGE, inner sep=8pt},
    channel/.style={draw, rectangle, rounded corners,
                    fill=yellow!40, minimum height=2.2cm, 
                    minimum width=3.8cm, align=center, 
                    font=\LARGE, inner sep=8pt},
    data/.style={draw, rectangle, rounded corners,
                 fill=pink!40, minimum height=2.2cm,
                 minimum width=3.4cm, align=center, 
                 font=\LARGE, inner sep=8pt},
    output/.style={draw, rectangle, rounded corners,
                   fill=cyan!40, minimum height=2.2cm,
                   minimum width=3.4cm, align=center, 
                   font=\LARGE, inner sep=8pt},
    arrow/.style={->, thick, line width=1.5pt}]

\node[data] (rawdata) at (0,0) 
    {\textbf{Raw}\\[5pt]\textbf{Data}};

\node[block] (foundation) at (4.5,0) 
    {\textbf{Foundation}\\[5pt]\textbf{Model}};

\node[block] (compression) at (9.2,0) 
    {\textbf{Semantic}\\[5pt]\textbf{Compression}};

\node[channel] (digchain) at (14,0) 
    {\textbf{Digital}\\[5pt]\textbf{Chain}};

\node[block] (reconstruction) at (18.8,0) 
    {\textbf{Embedding}\\[5pt]\textbf{Reconstruction}};

\node[block] (inference) at (23.5,0) 
    {\textbf{Inference}\\[5pt]\textbf{Model}};

\node[output] (output) at (27.8,0) 
    {\textbf{Inferred}\\[5pt]\textbf{Labels}};

\node[draw, dashed, rectangle, rounded corners,
      minimum height=3.4cm, minimum width=8.8cm,
      label={[font=\Large\bfseries]above:\textbf{Transmitter}}]
      at (11.5,0) {};

\node[draw, dashed, rectangle, rounded corners,
      minimum height=3.4cm, minimum width=8.9cm,
      label={[font=\Large\bfseries]above:\textbf{Receiver}}]
      at (20.95,0) {};

\draw[arrow] (rawdata) -- (foundation);

\draw[arrow] (foundation) -- (compression)
    node[midway, above=8pt, font=\Large] {$X$};

\draw[arrow] (compression) -- (digchain)
    node[midway, above=8pt, font=\Large] {$T$};

\draw[arrow] (digchain) -- (reconstruction)
    node[midway, above=8pt, font=\Large] {$\hat{B}$};

\draw[arrow] (reconstruction) -- (inference);

\draw[arrow] (inference) -- (output)
    node[midway, above=8pt, font=\Large] {$Y$};

\node[draw, dashed, rounded corners, fill=gray!10,
      font=\Large, align=center,
      minimum width=10cm, inner sep=10pt] (annotations)
    at (13.9,-2.9) 
    {$T \xrightarrow{\;Quantization\;} B 
    \xrightarrow{\;\mathrm{BSC}(p_{error})\;} 
    \hat{B}$};

\draw[arrow, dashed] (annotations) -- (digchain);

\end{tikzpicture}
}%
\captionsetup{justification=centering}
\caption{Schematic of the studied semantic communication system.}
\label{fig:semantic_comm_system}
\end{figure*}
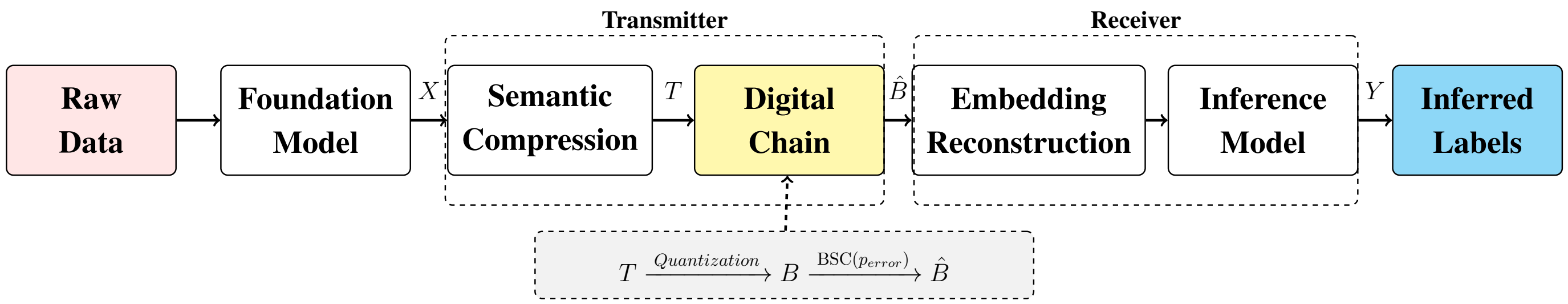

We consider a semantic communication system that operates on 
high-dimensional embeddings $X \in \mathbb{R}^n$ extracted from 
raw data using pre-trained foundation models. The system consists 
of three main components: a transmitter that compresses $X$ into 
$T \in \mathbb{R}^m$ where $m < n$, a digital communication chain 
that handles quantization and transmission, and a receiver that 
reconstructs embeddings from $\hat{B}$ for inference about target variable $Y$. 
A representation of the system is given in 
Fig.~\ref{fig:semantic_comm_system}.

For comparison purposes with~\cite{matos2025compacting}, we generate embeddings from raw text data by using the pre-trained model \textit{tomaarsen/mpnet-base-nli}\footnote{\url{https://huggingface.co/tomaarsen/mpnet-base-nli}}. This model outputs embeddings with $n=768$ dimensions, which are further compressed into $m \in \{32,128\}$.

\subsection{Compression Model}

Following the Gaussian Information Bottleneck framework, the transmitter implements:
\begin{equation}
    T = AX + Z,
    \label{eq:compression}
\end{equation}

where $A \in \mathbb{R}^{m \times n}$ is the compression matrix to be optimized, and $Z \sim \mathcal{N}(0, \Sigma_Z)$ represents controlled noise injection. The compressed representation then follows $T|X \sim \mathcal{N}(AX, \Sigma_Z)$.

\subsection{Quantization Scheme}

The continuous representation $T \in \mathbb{R}^m$ is quantized using K-means clustering with $L$ clusters. The quantization process involves:

\begin{enumerate}
    \item \textbf{Codebook generation}: K-means clustering is applied to a training set of compressed representations $\{T^{(i)}\}_{i=1}^{N_{train}}$ to obtain cluster centroids $\{\mathbf{c}_1, \mathbf{c}_2, \ldots, \mathbf{c}_L\}$,
    
    \item \textbf{Quantization mapping}: Each compressed representation is mapped to the nearest centroid:
    \begin{equation}
        B = \text{Quantize}(T) = \arg\min_{b} \|T - \mathbf{c}_{b}\|^2,
        \label{eq:quantization}
    \end{equation}
    
    After quantization, the original continuous representation $T$ can then be reconstructed as $\mathbf{c}_{B}$.
\end{enumerate}

The number of clusters $L$ determines the quantization rate of $R = \lceil \log_2 L \rceil$ bits per compressed embedding. The K-means quantization provides a practical balance between implementation simplicity and quantization performance.

\subsection{Channel Model}
The binary sequence $B$ is transmitted over a Binary Symmetric Channel (BSC) with a crossover probability $p_{error}$:

\begin{equation}
    P(\hat{B}_i \neq B_i) = p_{error}, \quad P(\hat{B}_i = B_i) = 1 - p_{error},
    \label{eq:channel}
\end{equation}

where $\hat{B} \in \{0,1\}^\ell$ is the received binary sequence.

\subsection{Distributional Assumptions}
\label{subsec:gaussian_assumps}

Our framework operates under relaxed Gaussian assumptions that are well-suited for high-dimensional text embeddings:

\begin{enumerate}
    \item \textbf{Conditional Gaussianity}: $X|Y=y \sim \mathcal{N}(\mu_{X|y}, \Sigma_{X|y})$ for each class $y$,
    
    \item \textbf{Discrete target}: $Y$ takes values in a finite set of classes, as in classification.
\end{enumerate}

No Gaussian assumption is imposed on the marginal $p(x)$. For discrete $Y$, it is the Gaussian mixture $p(x)=\sum_y p(y)\,\mathcal{N}(x;\mu_{X|y},\Sigma_{X|y})$, which may be multimodal and non-Gaussian, with the mixing distribution $p(y)$ left arbitrary. By contrast, classical GIB~\cite{chechik2005information} requires $(X,Y)$ to be jointly Gaussian. This makes $Y$ Gaussian, with $X$ and $Y$ related only linearly, and excludes discrete labels. The relaxation here is therefore from a single joint Gaussian to a mixture of Gaussian components indexed by the class label $Y$. Class-conditional Gaussianity is a modeling assumption for high-dimensional embeddings, and it enables analytical tractability while supporting tasks with discrete labels.

\section{Theoretical Foundation}
\label{sec:theory}
\subsection{Analytical Gradient Expressions}
\label{subsec:analytical_gradients}
In this subsection, we derive closed-form expressions for the gradients of the relaxed Gaussian information bottleneck objective under the proposed quantization and BSC model. These analytical expressions enable efficient optimization of the compression matrix $A$ while accounting for both quantization distortion and channel noise effects.

Under the data flow presented in Subsection~\ref{impl_challenges}, our goal is to optimize the objective function \eqref{eq:practical-IB-objective}, where the optimization is performed over the compression matrix $A$. From the linear transformation model presented in Subsection~\ref{subsec:GIB}, and the Gaussianity assumptions in Subsection~\ref{subsec:gaussian_assumps}, the conditional distributions of the compressed representation become:
\begin{align}
&T|X=x \sim \mathcal{N}(Ax, \Sigma_Z), \\
&T|Y=y \sim \mathcal{N}(A\mu_{X|y}, S),
\label{eq:t_given_y}
\end{align}
where $S = A\Sigma_{X|y}A^\top + \Sigma_Z$.

The quantization process assigns each realization $t$ to one of $L$ quantization regions $\mathcal{C}_b$, giving the quantization probability:
\begin{equation}
p(b|x) = \int_{\mathcal{C}_b} p(t|x) \, dt.
\label{eq:quantization_prob}
\end{equation}

The BSC introduces bit flips with probability $p_{error}$, characterized by:
\begin{equation}
p(\hat{b}|b) = \Pr(\hat{B}=\hat{b}|B=b),
\label{eq:bsc_prob}
\end{equation}
which depends only on the crossover probability $p_{error}$ and not on the compression matrix $A$.

Thus, the combined effect of quantization and channel noise yields:
\begin{align}
p(\hat{b}|x) &= \sum_{b} p(\hat{b}|b)p(b|x), \\
p(\hat{b}|y) &= \sum_{b} p(\hat{b}|b)p(b|y). \label{eq:combined_y}
\end{align}

From the definition of mutual information, the compression term is given by:
\begin{equation}
    I(X;\hat{B})=\int p(x)\sum_{\hat{b}}p(\hat{b}|x) \log\frac{p(\hat{b}|x)}{p(\hat{b})}dx,
\end{equation}

where $X$ is continuous and $\hat{B}$ is discrete. The dependence on $A$ is fully concentrated in $p(t|x)$. After its derivation, we find:
\begin{equation}
    \nabla_A I(X;\hat{B}) = \mathbb{E}_{p(x)} \left[ \sum_{\hat{b}} \nabla_A [p(\hat{b}|x)]\log\frac{p(\hat{b}|x)}{p(\hat{b})} \right].
\end{equation}

And equivalently,
\begin{equation}
    \nabla_A I(\hat{B};Y) = \mathbb{E}_{p(y)} \left[ \sum_{\hat{b}} \nabla_A [p(\hat{b}|y)]\log\frac{p(\hat{b}|y)}{p(\hat{b})} \right].
\end{equation}

Since $p(\hat{b}|b)$ is independent of $A$, we have:
\begin{align}
\nabla_A [p(\hat{b}|x)] &= \sum_{b} p(\hat{b}|b) \int_{\mathcal{C}_{b}} \nabla_A [p(t|x)] \, dt, \label{eq:grad_p_x} \\
\nabla_A [p(\hat{b}|y)] &= \sum_{b} p(\hat{b}|b) \int_{\mathcal{C}_{b}} \nabla_A [p(t|y)] \, dt, \label{eq:grad_p_y}
\end{align}
where the distributions $p(t|x)$ and $p(t|y)$ are Gaussian.\\

\subsubsection{Analyzing the Term $\nabla_A [p(t|x)]$}
Using the Gaussianity of $p(t|x)$, we find:
\begin{equation}
    p(t|x)=
\frac{1}{(2\pi)^{\frac{m}{2}}|\Sigma_Z|^{\frac{1}{2}}}
\exp\left(
-\frac{1}{2}(t-Ax)^\top \Sigma_Z^{-1}(t-Ax)
\right).
\end{equation}

From the identity $\nabla_A (x^\top V x)=2Vx x^\top$ for symmetric $V$~\cite{matrixcookbook20212}, we apply the log function and derive in relation to $A$ to obtain:
\begin{equation}
\nabla_A \log p(t|x)
=
\Sigma_Z^{-1}(t-Ax)x^\top.
\end{equation}

Writing $p(t|x) = \exp(\log(p(t|x))$ and using the chain rule, we finally have:
\begin{equation}
\nabla_A p(t|x)
=
p(t|x)\Sigma_Z^{-1}(t-Ax)x^\top.
\end{equation}

\subsubsection{Analyzing the Term $\nabla_A [p(t|y)]$}
Similarly to the previous case, the Gaussianity of $p(t|y)$ gives:
\begin{equation}
p(t|y)=
\frac{1}{(2\pi)^{\frac{m}{2}}|S|^{\frac{1}{2}}}
\exp\left(
-\frac{1}{2}(t-\mu_{T|y})^\top S^{-1}(t-\mu_{T|y})
\right),
\end{equation}
where $\mu_{T|y}=A\mu_{X|y}$.

Let us define $Q = S^{-1}(t-\mu_{T|y})(t-\mu_{T|y})^\top S^{-1} - S^{-1}$. Using both results $d\log|S|=\mathrm{tr}(S^{-1}dS)$, $dS^{-1}=-S^{-1}(dS)S^{-1}$~\cite{matrixcookbook20212}, and after the derivation provided in Appendix~\ref{append:condprobderivative}, we can find:

\begin{equation}
\nabla_A \log p(t|y)
=
S^{-1}(t-\mu_{T|y})\mu_{X|y}^\top
+
Q A \Sigma_{X|y}.
\end{equation}

Writing $p(t|y) = \exp(\log(p(t|y))$ and using the chain rule, we finally have:
\begin{equation}
\nabla_A p(t|y)
=
p(t|y)
\left[
S^{-1}(t-\mu_{T|y})\mu_{X|y}^\top
+
Q A \Sigma_{X|y}
\right].
\end{equation}

\subsubsection{Complete Gradient Expression}
Combining the above results, the gradient of the relaxed Gaussian information bottleneck objective becomes:
\begin{align}
\nabla_A \mathcal{L} &= \nabla_A I(X;\hat{B}) - \beta \nabla_A I(\hat{B};Y) \nonumber \\
&= \mathbb{E}_{p(x)} \left[ \sum_{\hat{b}}\sum_{b} p(\hat{b}|b) \int_{\mathcal{C}_b} \nabla_A[p(t|x)] \, dt \log\frac{p(\hat{b}|x)}{p(\hat{b})} \right] \nonumber \\
\quad &- \beta \mathbb{E}_{p(y)} \left[ \sum_{\hat{b}}\sum_{b} p(\hat{b}|b) \int_{\mathcal{C}_b} \nabla_A[p(t|y)] \, dt \log\frac{p(\hat{b}|y)}{p(\hat{b})} \right].
\label{eq:complete_gradient}
\end{align}

The optimal compression matrix is obtained by solving:
\begin{equation}
A^* = \arg\min_A \mathcal{L}(A),
\label{eq:optimal_A}
\end{equation}

which we can solve numerically by applying the previously described quantization scheme, and using the minibatch Stochastic Gradient Descent (SGD) approach. This analytical framework enables efficient gradient-based optimization while explicitly accounting for the statistical effects of both quantization and BSC-induced errors on the objective function.

\subsection{Theoretical Distortion-based Bound}
A fundamental challenge in our framework is connecting the continuous optimization objective involving $T$ with the practical discrete performance involving $\hat{B}$. While our analytical gradients optimize the compression matrix $A$ based on the continuous representation $T = AX + Z$, the actual system performance depends on the reconstructed representation from $\hat{B}$ obtained after quantization and channel transmission. We establish a theoretical bound that guarantees the continuous optimization translates to task performance from quantized representations, and propose to study the preservation of that property when including the channel effect.

\subsubsection{The Quantization Effect}
By only considering the quantization of $T$ to reconstruct discrete representations from $\hat{B}$, the simplified Markov chain becomes $Y \rightarrow X \rightarrow T \rightarrow B$.
Using the chain rule and the KL divergence-based definition of mutual information:
\begin{align}
    I(T;Y) = I(B;Y)& + I(T;Y|B) \nonumber \\
    I(T;Y) - I(B;Y) &= I(T;Y|B) \nonumber \\
    &= \mathbb{E}_{T,B}[D_{KL}(p(y|T,B)\,\big\|\, p(y|B))],
\end{align}
where the expectation is over the joint law of $(T,B)$. But because $B$ is produced from $T$ alone, $p(y|T,B) = p(y|T)$. Therefore,
\begin{equation}
    I(T;Y) - I(B;Y) = \mathbb{E}_{T,B}[D_{KL}(p(y|T)\,\big\|\, p(y|B))].
\end{equation}

Assume the log-likelihood function satisfies a Lipschitz continuity condition: there exists a constant $L_{Lip} \geq 0$ such that for all $y \in \mathcal{Y}$ label realizations and all $t, \hat{t} \in \mathbb{R}^m$,
\begin{equation}
    \left| \log p(y \mid t) - \log p(y \mid \hat{t}) \right| \leq L_{Lip} \, \|t - \hat{t}\|,
    \label{eq:lipschitz_condition}
\end{equation}

where $\hat{t}=\mathbf{c}_{b}$ is the reconstructed version of $t$, directly obtained from the quantization centroid assigned by $b$.

Then,
\begin{align}
D_{\mathrm{KL}}\!\left(p(y \mid t)\,\big\|\,p(y \mid \hat{t})\right) \nonumber
&= \int p(y \mid t)\,
\log \frac{p(y \mid t)}{p(y \mid \hat{t})}\, dy \\[4pt]
&= \int p(y \mid t)\, \nonumber
\big( \log p(y \mid t) \\& - \log p(y \mid \hat{t}) \big)\, dy \\
&\le \int p(y \mid t)\, L_{Lip} \|t - \hat{t}\| \, dy \nonumber \\
&= L_{Lip} \|t - \hat{t}\|.
\end{align}

Consequently,
\begin{equation}
0 \;\le\; I(T;Y) - I(B;Y)
\;\le\; L_{Lip}\, \mathbb{E}\!\left[\|T - \hat T\|\right].
\end{equation}

Using that square root is a concave function and $\|T - \hat{T}\|^2$ is a scalar random variable, we use the Jensen's inequality to obtain:
\begin{equation}
\sqrt{\mathbb{E}\left[\|T - \hat{T}\|^2\right]} \geq \mathbb{E}\left[\sqrt{\|T - \hat{T}\|^2}\right].
\end{equation}

Then, the mutual information between the continuous and reconstructed representations is bounded by:
\begin{equation}
    0 \leq I(T;Y) - I(B;Y) \leq L_{Lip} \, \sqrt{\mathbb{E}\left[\|T - \hat{T}\|^2\right]},
    \label{eq:mi_bound}
\end{equation}
which is exactly the objective function minimized by K-means, justifying its implementation in this work. Thus if the expected quantization distortion $\mathbb{E}\left[\|T - \hat{T}\|^2\right] \rightarrow 0$ by using K-means, then $I(T;Y) \rightarrow I(B;Y)$.

\subsubsection{Adding the BSC Distortion}
Considering the original data flow, the reconstructed version of $T$ is obtained from quantizing and subjecting their discrete representations to a BSC. The studied Markov chain then becomes $Y \rightarrow X \rightarrow T \rightarrow B \rightarrow \hat{B}$, where $B$ represents the intermediate variable obtained from the quantization of $T$, and $\hat{B}$ results from transmitting $B$ through BSC.

Using the chain rule twice for the concerned mutual information terms:
\begin{align}
    I(T;Y) - I(\hat{B};Y) &= I(T;Y) - I(B;Y) \\
                          &+ I(B;Y) - I(\hat{B};Y) \\
                          &= D_{Q} + D_{BSC},
\end{align}
where $D_{Q}$ and $D_{BSC}$ represent the distortion due to quantization and BSC effects individually.

Even if quantization distortion can be bounded and minimized through appropriate quantization schemes, the BSC introduces the additional distortion $D_{BSC} \geq 0$ that depends on the crossover probability $p_{error}$. Unlike quantization parameters, $p_{error}$ is typically unknown or variable in practical communication scenarios and cannot be directly optimized during compression design.

In this work, we propose to analyze the robustness of compression models against BSC-induced errors by studying how channel noise statistically affects gradient estimates during training. Rather than explicitly optimizing the information bottleneck-based objective as a function of $p_{error}$—which would require knowledge of the channel parameters—we develop training procedures that inherently take into account the channel conditions.

\section{Practical Algorithm Design}
\label{sec:algorithm}
The design primarily investigates whether using a relaxed Gaussian information bottleneck approach to compress and quantize general-purpose embeddings yields better performance than traditional quantization-based compression techniques, which fail to explicitly maintain information preservation between intermediate representations and application-level target variables. Specifically, we compare our results directly with those presented in~\cite{matos2025compacting} using the Product Quantization method, particularly focusing on results obtained without imposing any specific properties on the embedding model used to structure data representations—an approach that aligns with our work.

\subsection{Training the Compression Matrix}
Using the previous theoretical foundation to estimate gradients numerically, training optimizes the relaxed Gaussian information bottleneck objective. The compression matrix $\mathbf{A} \in \mathbb{R}^{m \times n}$ learns to compress semantic embeddings from dimension $n$ to $m$ while preserving task-relevant information under quantization and noisy channel conditions. Algorithm~\ref{alg:compression_training} details how the training matrices are obtained from numerical gradient estimation, where we define the equalities $\mathbf{S}_y = \mathbf{A}\boldsymbol{\Sigma}_{X|y}\mathbf{A}^T + \boldsymbol{\Sigma}_Z$, $\mathbf{v} = \mathbf{S}_y^{-1}(\mathbf{t} - \mathbf{A}\boldsymbol{\mu}_{X|y})$, and $\mathbf{P} = \mathbf{S}_y^{-1}[(\mathbf{t}-\mathbf{A}\boldsymbol{\mu}_{X|y})(\mathbf{t}-\mathbf{A}\boldsymbol{\mu}_{X|y})^T]\mathbf{S}_y^{-1} - \mathbf{S}_y^{-1}$.

\begin{algorithm}[h]
\small
\caption{Compression Matrix Training}
\label{alg:compression_training}
\begin{algorithmic}[1]
\REQUIRE Training embeddings $\mathbf{X} \in \mathbb{R}^{N_{train} \times n}$, training labels $\mathbf{Y}$, compressed dim $m$
\REQUIRE Parameters: $B,K,\eta, \beta, \epsilon_Z, n_{epochs}, L, p_{error}$
\ENSURE Compression matrix $\mathbf{A} \in \mathbb{R}^{m \times n}$

\STATE \textbf{Preprocessing:} PCA whitening $\mathbf{X}$; compute $\{\boldsymbol{\mu}_{X|y}, \boldsymbol{\Sigma}_{X|y}\}$ per class
\STATE \textbf{Initialize:} Random $\mathbf{A}$ (normalized), $\boldsymbol{\Sigma}_Z = \epsilon_Z \mathbf{I}_m$, Adam optimizer

\FOR{epoch $= 1$ to $n_{epochs}$}
\STATE Learn quantization codebook: $\mathcal{C} \leftarrow \text{KMeans}(\mathbf{X} \mathbf{A}^T, L)$
\FOR{each minibatch $\mathcal{B}$ of size $B$}
\STATE Sample: $\{\mathbf{t}_k \sim \mathcal{N}(\mathbf{A}\mathbf{x}_i, \boldsymbol{\Sigma}_Z)\}_{k=1}^K$ and $\{\mathbf{t}_k \sim \mathcal{N}(\mathbf{A}\boldsymbol{\mu}_{X|y_i}, \mathbf{S}_y)\}_{k=1}^K$
\STATE Quantize $t_k$ samples to nearest centroids; compute BSC probabilities $p(\hat{\mathbf{b}}|b)$
\STATE Compute importance weights: \\ $\omega_X, \omega_Y \leftarrow \sum_{\hat{b}} p(\hat{b}|b) \log \frac{p(\hat{b}|X,Y)}{p(\hat{b})}$
\STATE Compute analytic gradient: \\ $\nabla_\mathbf{A} \log p(\mathbf{t}|\mathbf{x}) = \boldsymbol{\Sigma}_Z^{-1}(\mathbf{t} - \mathbf{A}\mathbf{x})\mathbf{x}^T$
\STATE Compute analytic gradient: \quad\quad\quad\quad\quad\quad\quad \\ $\nabla_\mathbf{A} \log p(\mathbf{t}|\mathbf{y}) = \mathbf{v} \boldsymbol{\mu}_{X|y}^T + \mathbf{P} \mathbf{A} \boldsymbol{\Sigma}_{X|y}$
\STATE Aggregate: \\ $\hat{\nabla}_\mathbf{A} = \frac{1}{BK}\sum_{i,k} [\omega_X[i,k] \nabla_\mathbf{A} \log p(\mathbf{t}_k|\mathbf{x}_i)
- \beta \omega_Y[i,k] \nabla_\mathbf{A} \log p(\mathbf{t}_k|\mathbf{y}_i)]$
\STATE Update: $\mathbf{A} \leftarrow \text{Adam}(\mathbf{A}, \hat{\nabla}_\mathbf{A},\eta)$
\ENDFOR
\ENDFOR
\RETURN $\mathbf{A}$
\end{algorithmic}
\end{algorithm}

For Algorithm \ref{alg:compression_training}, the Principal Component Analysis (PCA) whitening centers the embeddings and rescales covariance eigendirections to reduce anisotropy and improve conditioning of the class-conditional statistics \cite{su2021whitening}. It is not meant to make the embeddings Gaussian: the theory requires only the class-conditional Gaussianity stated in Sec.~\ref{subsec:gaussian_assumps}. For generating the conditional distributions, we pre-compute the values $\{\mu_{X|y}, \Sigma_{X|y}\}$ per class $y$ from the available training embeddings. Also, we use the reparameterization trick with Cholesky decomposition for sampling from $T|X$ and $T|Y$ distributions, and Adam optimizer for handling gradient noise from Monte Carlo estimation.

\subsection{Evaluation Pipeline}
\subsubsection{Massive Text Embedding Benchmark}
Our evaluation leverages the comprehensive framework MTEB \cite{muennighoff2022mteb}  that provides standardized evaluation protocols for text embedding models across diverse tasks and datasets. MTEB enables a systematic assessment of how compression techniques affect embedding quality in practical application scenarios.

For our experiments, we select the aforementioned \textit{tomaarsen/mpnet-base-nli} as the base embedding model. This generalist architecture generates 768-dimensional vector representations and is designed to perform effectively across multiple natural language tasks without requiring task-specific adaptations. The choice of a general-purpose model aligns with our objective of evaluating compression strategies in scenarios where a single embedding representation must support various downstream applications within practical operated networks.

Also, we conduct our evaluation using the \textit{Banking77} dataset\footnotemark
\footnotetext{\url{https://huggingface.co/datasets/mteb/banking77}}, accessed through the MTEB framework for multilabel classification assessment. This corpus addresses intent recognition in the banking domain, containing customer service inquiries labeled according to their underlying purpose and organized in two splits: a training dataset with 9993 samples, and a testing dataset with 3080 samples.

The multilabel classification task associated with Banking77 provides an objective measure of compression impact: we can directly quantify whether compressed embeddings maintain sufficient semantic information to correctly classify customer intents across 77 different categories. This relatively large label space provides a more complex evaluation scenario where subtle information loss from compression could manifest as degraded classification accuracy, making it relevant for assessing the preservation capabilities of different compression approaches.

\subsubsection{Performance Metrics}
Our evaluation employs two complementary metrics that capture both network efficiency and application-level effectiveness.

\textbf{Communication Cost}: To quantify the network resources required for embedding transmission, we measure the bit budget per embedding $n_{bits}$—the total number of bits needed to communicate a single compressed representation.
    Under the PQ compression scheme proposed in~\cite{matos2025compacting}, the original 768-dimensional embedding is partitioned into $m$ equal-sized sub-vectors. Each sub-vector is then quantized using a codebook containing 2 centroids, allowing each quantized index to be encoded with a single bit. The resulting transmission cost is $n_{bits}=m$, where $m \in \{1,2,4,8,16,32,64,128,256,768\}$ represents the number of sub-vector partitions created.
    For our proposed relaxed GIB-based compression framework, the bit budget per embedding is directly defined by the quantization rate applied on K-means, and gives then $n_{bits} = R$ as the number of bits to quantize the new embedding spaces obtained from linear compression and defined by $m \in \{32,128\}$ dimensions. In this work, the quantization schemes are designed along the range $L \in \{2,4,32,64,128,256,512\}$ clusters, which immediately gives $R \in \{1,2,5,6,7,8,9\}$ bits to represent each quantization index.

\textbf{Task Performance Quality}: To measure application-level effectiveness, we compute classification accuracy according to MTEB evaluation standards. This metric directly indicates whether compressed embeddings retain sufficient semantic information for accurate intent recognition, and is computed as $\mathrm{Acc} = \frac{CP}{TP},$
where $CP$ stands for the number of test samples correctly predicted, and $TP$ stands for the total number of samples in the test set.

\subsubsection{The Evaluation Stage}
Following MTEB standards, we employ logistic regression as the task-specific classifier for intent inference. As the implemented evaluation stage, the classifier training and testing phases are structured as follows:

\textbf{Training phase}: The classifier is trained using labeled embeddings generated by the base model. Adhering to MTEB's standardized configuration, we sample $8$ balanced examples per intent class, resulting in a training set of $616$ sentences for Banking77 (i.e., $8$ samples $\times$ $77$ classes).
For PQ compression, a single classifier $C_{PQ}$ is trained on the 768-dimensional embeddings without applying compression during the training phase~\cite{matos2025compacting}.
For our proposed framework, the task-specific model receives compressed versions of the original embeddings after reception, and then a dedicated classifier model $C_{m}$ is trained for each $m \in \{32,128\}$. For this reason, training embeddings are first transformed with PCA whitening and then compressed with the compression matrices previously computed.

\textbf{Testing phase}: After each training round, the testing phase is also defined differently depending on the evaluated strategy. 
Following the evaluation workflow in~\cite{matos2025compacting}, PQ compression compacts the 768-dimensional test embeddings into multiple sub-vectors to compose PQ codes, which are transmitted through an identity channel and used at reception to reconstruct the test vectors at their original size.

For the relaxed GIB-based compression, given the compression matrix $\mathbf{A}$, the channel condition $p_{error}$, and the fixed quantization codebook $\mathcal{C}$ obtained with K-means, the testing phase is defined as in Algorithm \ref{alg:evaluation}.

\begin{algorithm}[ht]
\small
\caption{Testing phase for the proposed framework}
\label{alg:evaluation}
\begin{algorithmic}[1]
\REQUIRE Test sentences $\mathbf{X}_{\text{test}} \in \mathbb{R}^{N_{test} \times n}$, test labels $\mathbf{Y}_{\text{test}}$, trained $\mathbf{A} \in \mathbb{R}^{m \times n}$, channel model BSC, quantization codebook $\mathcal{C}$
\ENSURE Classification accuracy
\STATE Extract embeddings; apply PCA whitening
\STATE Compress: $\mathbf{T} = \mathbf{X}_{\text{test}} \mathbf{A}^T$
\STATE Quantize with $\mathcal{C}$: assign to nearest centroids $\rightarrow$ indices $\mathbf{B}$
\STATE Simulate channel: $\hat{\mathbf{B}} = \text{BSC}(\mathbf{B}, p_{error})$
\STATE Reconstruct $\mathbf{T}$ as $\hat{\mathbf{T}}$ from received $\hat{\mathbf{B}}$
\STATE Feed the classifier $C_{m}$ with $\hat{\mathbf{T}}$ for inference to get $\mathbf{Y}_{\text{test}}$
\STATE Compute the accuracy $\mathrm{Acc}$
\RETURN Accuracy $\mathrm{Acc}$
\end{algorithmic}
\end{algorithm}

Similarly to~\cite{matos2025compacting}, we denote one training and testing phase as one MTEB experiment, and we run 100 MTEB experiments for 100 distinct K-means initializations, ensuring random training sample selection across 10k different MTEB experiments. Additionally, the quantization codebooks are equally obtained from the entire training dataset.

\section{Experimental Results and Discussion}
\label{sec:experiments}

\begin{figure}
    \centering
    \includegraphics[width=3.5in]{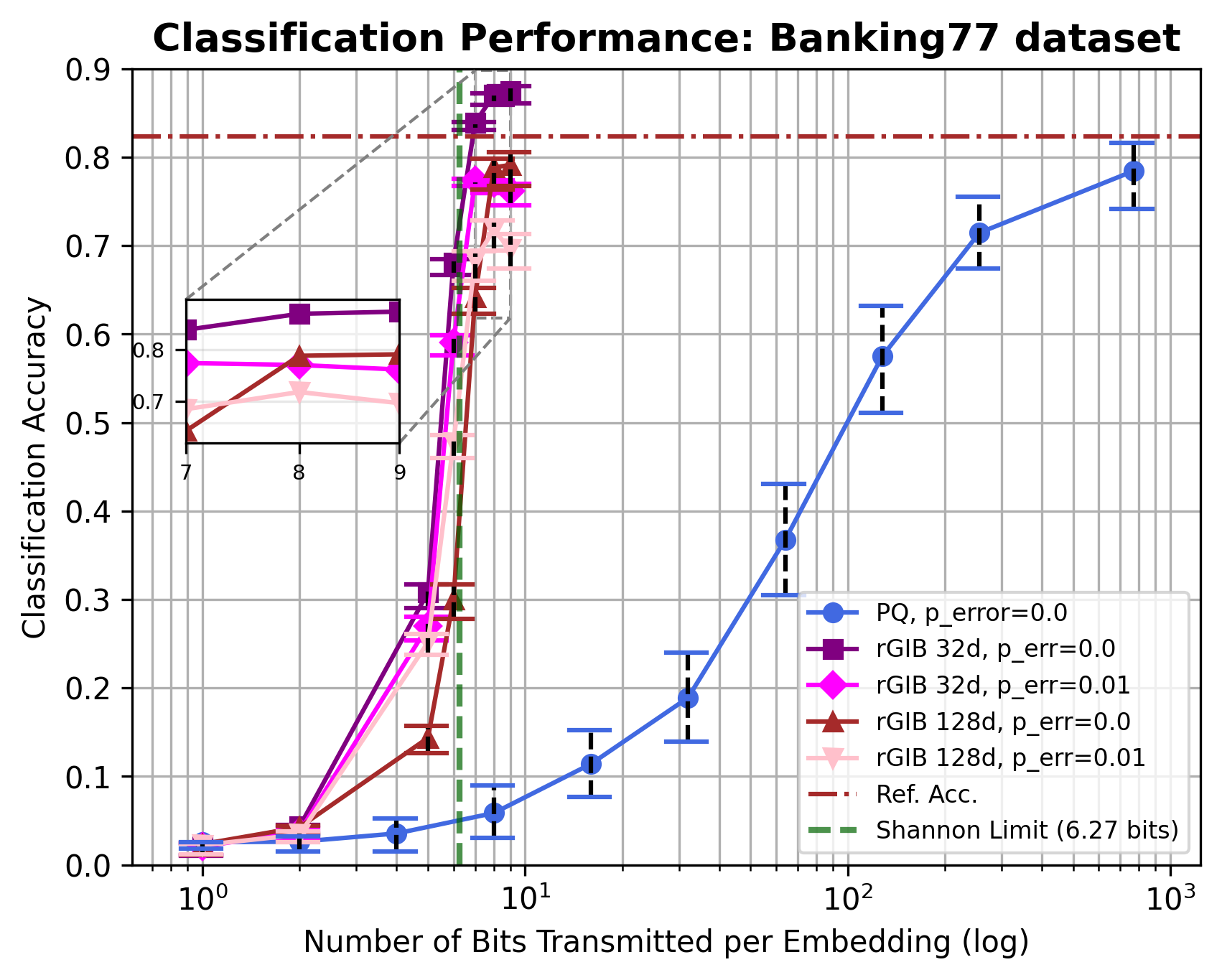}
    \caption{Accuracy performances against bit budget per embedding over 10k MTEB experiments for Banking77. The rGIB curves represent the relaxed GIB-based compression performances, while PQ curve indicates performances of the compression method implemented in~\cite{matos2025compacting}.}
    \label{fig:acc_x_bits-banking77}
\end{figure}

Table~\ref{tab:training_params} summarizes the main experimental parameters to train matrices in an offline stage for each compression configuration in Fig.~\ref{fig:acc_x_bits-banking77}, whose task performances are obtained in an online evaluation stage.

\begin{table}[t]
\centering
\caption{Compression matrix training parameters.}
\label{tab:training_params}
\begin{tabular}{lcc}
\toprule
\textbf{Parameter} & \textbf{Symbol} & \textbf{Value} \\
\midrule
Batch size              & $B$                        & 64             \\
Monte Carlo samples per instance    & $K$                        & 120            \\
Learning rate           & $\eta$                     & 0.001          \\
Beta coefficient        & $\beta$                    & 1.0            \\
Noise variance          & $\epsilon_Z$               & 0.01           \\
Training epochs         & $n_{epochs}$                 & 500          \\
\bottomrule
\end{tabular}
\end{table}

To validate the Lipschitz continuity assumption underlying our information-theoretic
bound, we numerically evaluate the empirical constant $\hat{L}_\mathrm{emp} = \max_{i,y} |\log p(y|t_i) - \log p(y|\hat{t}_i)| \,/\, \|t_i - \hat{t}_i\|$ across all test samples and classes, using the same logistic regression classifier employed by the MTEB evaluation pipeline.
MTEB experiments are repeated 100 times with independent random subsampling, fixed codebook designed from the entire training dataset, and $p_{error}=0.0$. We then report the maximum $\hat{L}_\mathrm{emp}$ over all 100 MTEB experiments, and their respective average accuracy for both non-quantized ($t$) and quantized ($\hat{t}$) compressed embeddings. Results for compressed dimensions $m \in \{32, 128\}$ and codebook sizes
$L \in \{2, \ldots, 4096\}$ are shown in Table~\ref{tab:lipschitz_verification}.
In every configuration, $\hat{L}_\mathrm{emp}$ is finite and bounded---it does not
exceed $4.01$ for $m=32$ and $1.66$ for $m=128$---confirming that the Lipschitz
assumption holds empirically across all evaluated settings.

\begin{table}[t]
  \centering
  \caption{Empirical verification of the Lipschitz continuity assumption
           on Banking77.
           $m$: compressed dimension; $L$: codebook size;
           $\hat{L}_\mathrm{emp}$: empirical Lipschitz constant. All accuracy values in \%, averaged over 100 MTEB experiments 
           with standard MTEB training.}
  \label{tab:lipschitz_verification}
  \renewcommand{\arraystretch}{1.1}
  \setlength{\tabcolsep}{3pt}
  \scriptsize
  \begin{tabular}{r r S[table-format=2.1] S[table-format=2.1] S[table-format=2.4] S[table-format=1.2]}
    \toprule
    $m$ & $L$
      & {$\mathrm{Acc}(t)$\,}
      & {$\mathrm{Acc}(\hat{t})$\,}
      & {$\mathbb{E}[\|t_i-\hat{t}_i\|]$}
      & {$\hat{L}_{\mathrm{emp}}$} \\
    \midrule
    \multirow{10}{*}{32} & 2 & 14.4 & 2.0 & 5.6365 & 2.88 \\
     & 4 & 15.0 & 4.3 & 5.8089 & 3.31 \\
     & 32 & 43.6 & 30.1 & 5.1920 & 4.01 \\
     & 64 & 76.5 & 67.5 & 4.5663 & 3.34 \\
     & 128 & 83.3 & 83.1 & 4.1703 & 2.88 \\
     & 256 & 83.7 & 87.0 & 3.8219 & 2.91 \\
     & 512 & 84.6 & 87.3 & 3.4899 & 2.97 \\
     & 1024 & 84.9 & 87.0 & 3.1077 & 2.88 \\
     & 2048 & 85.0 & 85.9 & 2.8028 & 3.02 \\
     & 4096 & 83.7 & 84.4 & 2.5011 & 3.12 \\
    \midrule
    \multirow{10}{*}{128} & 2 & 39.1 & 2.3 & 13.6441 & 1.40 \\
     & 4 & 39.8 & 4.3 & 13.6907 & 1.28 \\
     & 32 & 50.7 & 14.6 & 11.2827 & 1.66 \\
     & 64 & 67.7 & 26.6 & 10.9195 & 1.56 \\
     & 128 & 77.3 & 65.2 & 11.0526 & 1.48 \\
     & 256 & 77.2 & 79.5 & 10.8586 & 1.44 \\
     & 512 & 76.7 & 78.5 & 10.5945 & 1.44 \\
     & 1024 & 74.8 & 77.0 & 10.2721 & 1.44 \\
     & 2048 & 73.8 & 75.9 & 10.0422 & 1.41 \\
     & 4096 & 70.6 & 72.3 & 9.8845 & 1.43 \\
    \bottomrule
  \end{tabular}
\end{table}

\begin{table}[t]
  \caption{Effect of increasing the number of Monte Carlo samples $K$ 
during compression matrix $A$ training, for 
$L \in \{1024, 2048, 4096\}$ and $m \in \{32, 128\}$, 
extending the evaluation beyond the practical operating 
range shown in Fig.~\ref{fig:acc_x_bits-banking77}.
    All accuracy values in \%, averaged over 100 MTEB experiments 
    with standard MTEB training.}
  \label{tab:ksamples_comparison}
  \centering
  \renewcommand{\arraystretch}{1.05}
  \setlength{\tabcolsep}{3pt}
  \scriptsize
  \begin{tabular}{
      c
      r
      S[table-format=2.1]
      S[table-format=2.1]
      S[table-format=2.1]
      S[table-format=2.1]
      S[table-format=2.1]
      S[table-format=2.1]}
    \toprule
    & & \multicolumn{2}{c}{$K=120$}
      & \multicolumn{2}{c}{$K=220$}
      & \multicolumn{2}{c}{$K=1024$} \\
    \cmidrule(lr){3-4}
    \cmidrule(lr){5-6}
    \cmidrule(lr){7-8}
    {$m$} & {$L$}
      & {$\mathrm{Acc}(t)$}
      & {$\mathrm{Acc}(\hat{t})$}
      & {$\mathrm{Acc}(t)$}
      & {$\mathrm{Acc}(\hat{t})$}
      & {$\mathrm{Acc}(t)$}
      & {$\mathrm{Acc}(\hat{t})$} \\
    \midrule
    \multirow{3}{*}{32}
      & 1024 & 84.9 & 87.0 & 85.2 & 87.5 & 86.3 & 87.6 \\
      & 2048 & 85.0 & 85.9 & 86.1 & 86.8 & 86.6 & 87.3 \\
      & 4096 & 83.7 & 84.4 & 84.2 & 84.9 & 83.2 & 84.5 \\
    \midrule
    \multirow{3}{*}{128}
      & 1024 & 74.8 & 77.0 & 78.0 & 80.1 & 82.5 & 84.1 \\
      & 2048 & 73.8 & 75.9 & 76.7 & 79.0 & 82.3 & 83.6 \\
      & 4096 & 70.6 & 72.3 & 76.4 & 76.8 & 82.3 & 82.6 \\
    \bottomrule
  \end{tabular}
\end{table}

\begin{table}[t]
  \caption{Classifier evaluated on quantized test vs.\ quantized
    training embeddings with standard MTEB training (616 samples, at left) and full data training (all 9993 samples, at right) for fixed anchor matrix $A_L$.
    $\mathrm{Acc}(t)$: unquantized baseline accuracy on 
    fixed anchor $A$. All accuracy values in \%, averaged over 100 MTEB experiments.}
  \label{tab:fulldata_traineval}
  \centering
  \renewcommand{\arraystretch}{1.05}
  \setlength{\tabcolsep}{3pt}
  \scriptsize
  \begin{tabular}{
      c
      S[table-format=2.1]
      S[table-format=2.1]
      S[table-format=1.4]
      S[table-format=2.1]
      S[table-format=1.4]}
    \toprule
    \multicolumn{6}{c}{$m=32$, anchor $A_{L=512}$,\quad
      $\mathrm{Acc}(t)$: test\;86.1\%,\ train\;99.0\%} \\
    \cmidrule(lr){1-6}
    & {616 samp.}
      & \multicolumn{2}{c}{Test set eval., full train}
      & \multicolumn{2}{c}{Train set eval., full train} \\
    \cmidrule(lr){2-2}
    \cmidrule(lr){3-4}
    \cmidrule(lr){5-6}
    {$L$}
      & {$\mathrm{Acc}(\hat{t})$\,}
      & {$\mathrm{Acc}(\hat{t})$\,}
      & {$\mathbb{E}{\left[\left\|t_i-\hat{t}_i\right\|\right]}$}
      & {$\mathrm{Acc}(\hat{t})$\,}
      & {$\mathbb{E}{\left[\left\|t_i-\hat{t}_i\right\|\right]}$} \\
    \midrule
    2    &  2.5 &  2.5 & 7.0198 &  2.5 & 6.9902 \\
    4    &  5.0 &  4.6 & 6.3573 &  5.2 & 6.3319 \\
    32   & 36.8 & 37.0 & 4.8757 & 44.0 & 4.7325 \\
    64   & 70.4 & 70.2 & 4.2493 & 79.4 & 4.0398 \\
    128  & 85.4 & 85.4 & 3.8671 & 92.8 & 3.6100 \\
    256  & 86.7 & 86.8 & 3.6430 & 94.7 & 3.2956 \\
    512  & 87.3 & 87.5 & 3.4899 & 94.8 & 3.0452 \\
    1024 & 87.3 & 87.8 & 3.4003 & 94.4 & 2.7859 \\
    2048 & 86.7 & 87.5 & 3.3191 & 95.5 & 2.3546 \\
    4096 & 85.9 & 87.2 & 3.2426 & 97.5 & 1.6137 \\
    8192 & {--} & 87.0 & 3.1448 & 98.9 & 0.3747 \\
    9993 & {--} & 87.2 & 3.1149 & 99.0 & 0.0000 \\
    \bottomrule
  \end{tabular}

  \vspace{4pt}

  \begin{tabular}{
      c
      S[table-format=2.1]
      S[table-format=2.1]
      S[table-format=2.4]
      S[table-format=2.1]
      S[table-format=2.4]}
    \toprule
    \multicolumn{6}{c}{$m=128$, anchor $A_{L=256}$,\quad
      $\mathrm{Acc}(t)$: test\;85.2\%,\ train\;100.0\%} \\
    \cmidrule(lr){1-6}
    & {616 samp.}
      & \multicolumn{2}{c}{Test set eval., full train}
      & \multicolumn{2}{c}{Train set eval., full train} \\
    \cmidrule(lr){2-2}
    \cmidrule(lr){3-4}
    \cmidrule(lr){5-6}
    {$L$}
      & {$\mathrm{Acc}(\hat{t})$\,}
      & {$\mathrm{Acc}(\hat{t})$\,}
      & {$\mathbb{E}{\left[\left\|t_i-\hat{t}_i\right\|\right]}$}
      & {$\mathrm{Acc}(\hat{t})$\,}
      & {$\mathbb{E}{\left[\left\|t_i-\hat{t}_i\right\|\right]}$} \\
    \midrule
    2    &  1.7 &  1.6 & 15.1467 &  1.7 & 15.0634 \\
    4    &  3.7 &  3.8 & 14.6348 &  4.4 & 14.5557 \\
    32   & 26.8 & 27.1 & 12.3500 & 30.1 & 12.3196 \\
    64   & 53.6 & 53.7 & 11.8719 & 58.8 & 11.6543 \\
    128  & 74.8 & 75.1 & 11.3390 & 81.7 & 10.9941 \\
    256  & 79.5 & 80.3 & 10.8586 & 86.5 & 10.3479 \\
    512  & 79.7 & 81.0 & 10.5349 & 85.7 &  9.6923 \\
    1024 & 79.5 & 81.4 & 10.3108 & 85.9 &  8.7970 \\
    2048 & 78.6 & 80.9 & 10.0370 & 87.9 &  7.3919 \\
    4096 & 77.2 & 80.4 &  9.8560 & 94.1 &  5.0747 \\
    8192 & {--} & 81.2 &  9.6261 & 99.6 &  1.1864 \\
    9993 & {--} & 81.2 &  9.5320 &100.0 &  0.0000 \\
    \bottomrule
  \end{tabular}
\end{table}

\begin{figure}
    \centering
    \includegraphics[scale=0.355]{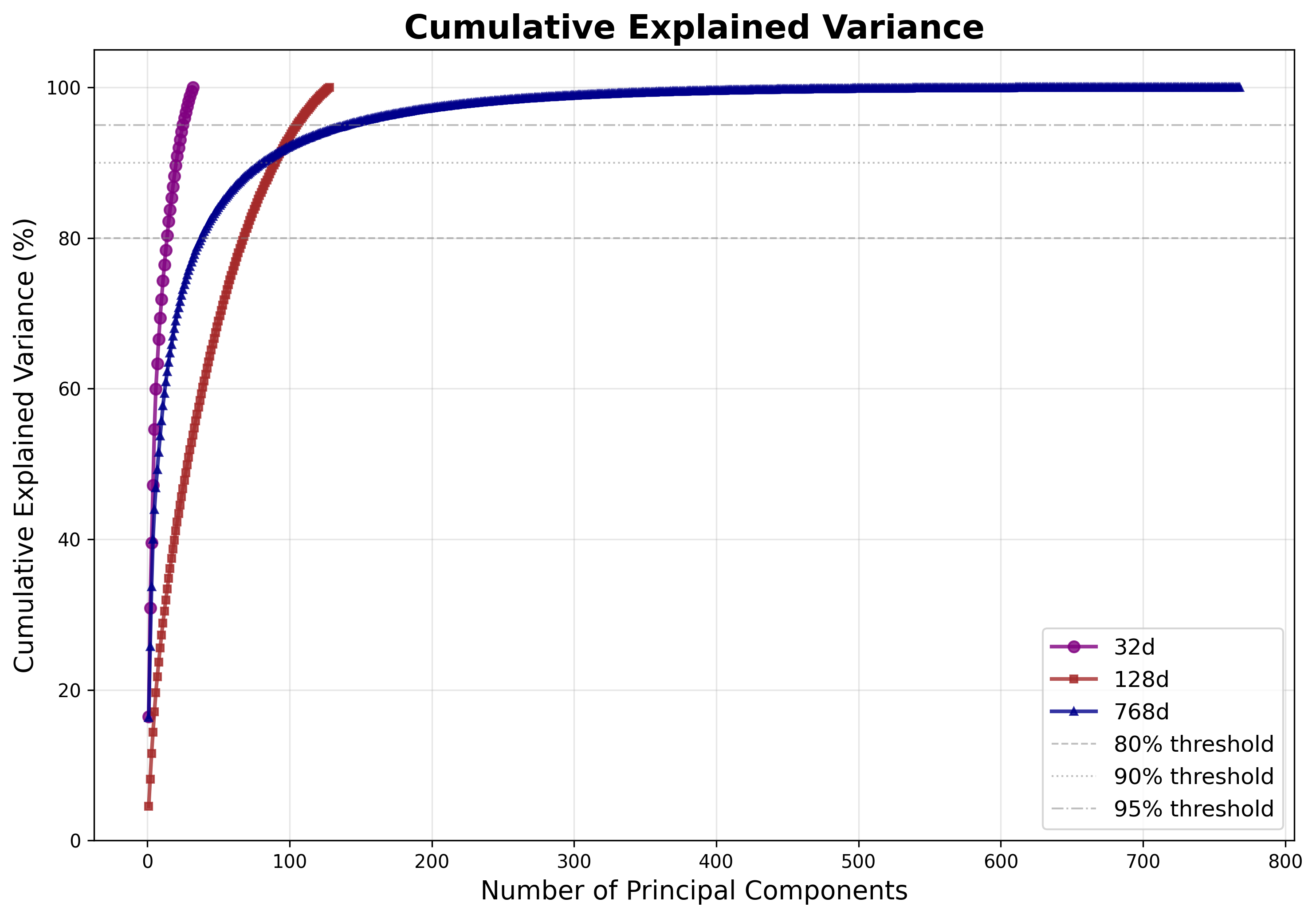}
    \caption{Cumulative explained variance across principal components for the three dimensional spaces evaluated.}
    \label{fig:cumul_exp_var}
\end{figure}

Fig.~\ref{fig:acc_x_bits-banking77} presents the average classification accuracy over 10k different MTEB experiments along the set of possible values of bit budget per embedding $n_{bits}$ for both PQ and relaxed GIB-based compression approaches, where the proposed scheme is referred to as \textit{rGIB} for simplicity. Illustrated as a horizontal line, the reference accuracy $\mathrm{Acc}_{ref}$ is defined as the maximum accuracy value obtained over 10k MTEB experiments using the full-dimensional embeddings without any particular processing, and presented in Fig.~\ref{fig:acc_x_bits-banking77} as $\mathrm{Acc}_{ref} = 0.8237$. Stated as the theoretical limit to perform lossless compression, the Shannon entropy is presented as the vertical green line and is computed over the label distribution on testing dataset. The relaxed GIB-based compression curves are presented for both compressed dimensions $m \in \{32, 128\}$, and BSC crossover probability $p_{error} \in \{0.0, 0.01\}$. Vertical dashed lines with colored boundaries indicate the maximum and minimum accuracy values observed across the 10,000 independent MTEB experiments for each compression configuration.

Fig.~\ref{fig:cumul_exp_var} shows the cumulative explained variance as a function of the number of principal components that allow to aggregate and describe the data distributions within each of the 32, 128 and 768-dimensional spaces. Horizontal lines represent how many principal components are needed to achieve $80\%$, $90\%$, and $95\%$ of explained variance on observed data.

Fig.~\ref{fig:acc_x_bits-banking77} demonstrates that the proposed relaxed GIB-based approach consistently outperforms PQ schemes across various compression configurations, delivering significantly superior accuracy even at low bit budgets per embedding. This finding reveals that the proposed compression can effectively preserve the essential information structure in general-purpose embeddings that are not specifically fine-tuned to perform a given task. The relaxed GIB-based method performs a specialized extraction of the most relevant features for performing the target task and adapts the intermediate task-oriented representations to better reflect the target distributions through the design of their quantized versions. Notably, similar classification accuracy can be obtained by the proposed method at 9 bits (i.e., the curve with 128 dimensions and $p_{error}=0.0$) and PQ at 768 bits, which demonstrates an efficiency gain of around \textbf{85 times} in favor of the proposed task-oriented compression compared to the task-agnostic PQ strategy. Importantly, this efficiency gain is achieved without modifying the upstream embedding model or its training procedure: 
the proposed linear compression operates as a task-specific post-processing 
step applied to the standard embedding output, preserving the 
foundation model as a universal, task-agnostic data encoder while 
adapting its output representations to better reflect the target 
task structure. Additionally, we observe that for the Banking77 intent classification task, the proposed method can achieve high accuracy performances from quantization-based coding schemes that approach the Shannon limit defined by the uniform label distribution in the testing dataset. Relative to the computed Shannon entropy, quantization schemes compressing more than the theoretical bound show highly degraded performances even if performed in compressed spaces composed of specialized feature extraction; while those respecting the compression limit can be very close to the referred bound and yet provide remarkable task performances for both reduced spaces.

While Fig.~\ref{fig:acc_x_bits-banking77} presents results 
up to $L=512$ (i.e., $n_{bits}=9$), which corresponds to the 
practical operating range of the proposed framework under the 
default $K=120$ Monte Carlo samples, Table~\ref{tab:ksamples_comparison} 
further characterizes the behavior at larger codebook sizes 
$L \in \{1024, 2048, 4096\}$.
For the proposed \textit{rGIB} framework, beyond $L=512$, the default number of Monte Carlo samples $K=120$ starts to become insufficient to well distribute the mass probabilities across all possible $L$ clusters during compression matrix training, which impacts the gradient estimation quality and is translated into degradation on accuracy values during inference. This effect is particularly evaluated and validated in Table~\ref{tab:ksamples_comparison}, where increasing the value of $K$ not only provides better accuracy values for both compressed spaces, but especially reduces the gradient distortion gap between successive values of large $L$ when $K$ is large enough to capture properly all cluster probabilities in quantizing $T$. This behavior is especially prominent for the 128-dimensional case due to the increasing dimensionality complexity.

Additionally, Table~\ref{tab:fulldata_traineval} also suggests that there is an intrinsic generalization gap - between training centroids and test embeddings - degrading accuracy performances at large $L$ under standard MTEB subsampling. Even in the case when we fix anchor matrices $A_{L}$ for compressing embeddings, we observe a particular loss of accuracy for large $L$ when MTEB classifiers have lower expressivity (i.e. by only training over a small subset of 616 samples). This behavior changes when classifiers have enough samples to better generalize over the test set, which is reconstructed from training centroids, achieving well defined accuracy plateaus. This case also reveals a fundamental reconstruction distortion that persists regardless of quantization refinement (e.g. the limit case of 9993 centroids representing each training embedding) due to the generalization gap. Also, the train-set evaluation side confirms the direct correspondence between quantization distortion and accuracy performance when no generalization gap is present.

An interesting observation in Fig.~\ref{fig:acc_x_bits-banking77} is that the MTEB classifier can better discriminate the label distribution from the quantization of the 32-dimensional space, supporting the hypothesis that there is a specific informational structure defined by a reduced set of features which are supposed to be the most relevant and suitable to well describe the task objective during classifier training. In this perspective, the 768-dimensional original space contains all the available information structured by the base model; however, it is organized as data representations that are too complex to capture the target label distribution. This complexity persists in the 128-dimensional embeddings in relation to the lower dimensional case, still retaining substantial feature information for representing labels to the classifier; nevertheless, the compression framework seems to allow sufficiently accurate quantization designs to maintain average accuracy approximating the uncompressed baseline. Besides, Fig.~\ref{fig:cumul_exp_var} illustrates that both compressed spaces have explained variance increasing more gradually in relation to their total number of dimensions, while the original embedding space concentrates the most part of its data distribution along a fewer number of principal components (i.e., $95\%$ of explained variance across 141 principal directions, against 25 and 106 principal components for the 32 and 128-dimensional cases, respectively). This suggests that, as explicitly intended by the optimization of the IB objective function \eqref{eq:practical-IB-objective}, the compression process eliminates redundant dimensions while preserving the label discrimination property in a more evenly distributed manner, potentially explaining the improved accuracy performance observed with 32-dimensional vectors, and similar classification performance of the 128-dimensional case in relation to reference value $\mathrm{Acc}_{ref}$. This observation highlights the inherent redundancy present in the general-purpose embedding model, which tend to increase representational complexity without necessarily adding task-relevant information. It further underscores the value of our proposed framework in constructing more task-informative spaces through specialized feature extraction.

As a result of the observed complexity, the 128-dimensional case (i.e., both curves with $p_{error} \in \{0.0, 0.01\}$) also presents a particular average accuracy distinction around the Shannon limit, where the quantization designs do not have enough centroids to cluster properly the label information, which changes when we perform finer space quantizations. The high dimensionality of these representations, allied to high quantization distortions on K-means as a hard clustering technique, can affect the convergence of SGD optimization and generate performance degradation even with no channel errors. A different behavior is shown, however, in the 32-dimensional cases, where a more stable performance pattern is observed over all compression configurations, including the ones at higher quantization distortion levels. Nevertheless, all relaxed GIB-based results, for both crossover probabilities $p_{error}$, extensively outperform the task-agnostic PQ compression with no channel errors, and reveal the potential of better designing task-oriented compression when the source data is intended to be represented as general-purpose vectors with no particular properties. Also, for all \textit{rGIB} curves and after the Shannon limit, we observe more explicitly the channel effect on curves with $p_{error}=0.01$, where increasing the number of bits also leads to higher reconstruction error probabilities for the quantized representations.

\section{Conclusion}
\label{sec:conclusion}
This paper presented a practical compression framework for semantic communication over wireless channels, grounded in a relaxed Gaussian Information Bottleneck formulation. By demonstrating that the framework requires only class-conditional Gaussianity of the source given the (discrete) target, with no Gaussian assumption on the marginal source distribution, we extended the applicability of GIB-based compression to classification tasks with discrete labels, significantly broadening its practical relevance. We established a rigorous distortion-based theoretical foundation that bounds the gap between continuous GIB optimization and its practical discrete counterpart under quantization, providing formal guarantees on task performance at reception. Building on this foundation, we derived closed-form analytical gradient expressions for the discrete GIB objective, enabling efficient optimization of the compression matrix $\mathbf{A}$. These theoretical contributions were integrated into a complete practical compression framework combining continuous matrix optimization with K-Means-based quantization design and iterative codebook updates. Comprehensive experiments on the Banking77 dataset demonstrated the effectiveness of the proposed approach against the task-agnostic PQ compression and reference baseline across multiple compression configurations, confirming that the framework achieves high task performance and remarkable efficiency under practical network scenarios. Particularly, the proposed framework demonstrated its ability to serve as a robust and specialized feature extraction method, identifying the most relevant task-oriented features that describe the target information from general-purpose representations without any particular property.

\appendix[The conditional log-probability $\log p(t|y)$ gradient]
\label{append:condprobderivative}
From the Gaussianity assumptions, we express $p(t|y)$ as
\begin{equation}
p(t|y)=
\frac{1}{(2\pi)^{\frac{m}{2}}|S|^{\frac{1}{2}}}
\exp\left(
-\frac{1}{2}(t-\mu_{T|y})^\top S^{-1}(t-\mu_{T|y})
\right),
\end{equation}
where $\mu_{T|y}=A\mu_{X|y}$, and $S = A\Sigma_{X|y}A^\top + \Sigma_Z$. Then,
\begin{align}
\nabla_A \log p(t|y) &= -\frac{1}{2} \left[ \nabla_A \log|S| \right. \nonumber \\
&\left. \quad + \nabla_A\left[ (t-\mu_{T|y})^\top S^{-1} (t-\mu_{T|y}) \right] \right].
\end{align}

Analyzing the expression over infinitesimal variations, we have
\begin{align}
    d \log p(t|y) &= -\frac{1}{2} \left[ d \log|S| \right. \nonumber \\
    &\left. \quad + \, d\left[ (t-\mu_{T|y})^\top S^{-1} (t-\mu_{T|y}) \right] \right],
\end{align}

where we define:
\begin{align}
    I   &= d \log|S|, \nonumber \\
    II  &= d\left[ (t-\mu_{T|y})^\top S^{-1} (t-\mu_{T|y}) \right].
\end{align}

From \cite{matrixcookbook20212}, we use the identities $d\log|S|=\mathrm{tr}(S^{-1}dS)$, $dS^{-1}=-S^{-1}(dS)S^{-1}$ to find
\begin{align}
    I   &= \mathrm{tr}(S^{-1}dS), \nonumber \\
    II  &= -\mu_{X|y}^\top(dA)^\top S^{-1}(t-\mu_{T|y}) \nonumber \\
    &-(t-\mu_{T|y})^\top S^{-1} (dA) \mu_{X|y} \nonumber \\
    &-(t-\mu_{T|y})^\top S^{-1} (dS) S^{-1} (t-\mu_{T|y}).
\end{align}

Using the product identity $a^\top Bb = b^\top B a$ for a symmetric matrix $B$ and vectors $a, b$ with proper dimensions, we can use that $S^{-1}$ is a symmetric matrix to simplify the second equation and obtain $II = -2(t-\mu_{T|y})^\top S^{-1} (dA) \mu_{X|y} -(t-\mu_{T|y})^\top S^{-1} (dS) S^{-1} (t-\mu_{T|y})$. Returning to the entire expression of $d \log p(t|y)$, we have
\begin{align}
    d \log p(t|y) &= (t-\mu_{T|y})^\top S^{-1} (dA) \mu_{X|y} \nonumber \\
    &+ \frac{1}{2} [ (t-\mu_{T|y})^\top S^{-1} (dS) S^{-1} (t-\mu_{T|y}) \nonumber \\
    &\phantom{{}+\frac{1}{2}[} - \mathrm{tr}(S^{-1}dS) ].
\end{align}

Applying the trace linearity, the identity $a^\top b = \mathrm{tr}(ba^\top)$ for $a,b$ real column vectors, and deriving $dS = (dA)\Sigma_{X|y}A^\top + A\Sigma_{X|y}(dA)^\top$, we can find
\begin{align}
    d \log p(t|y) &= (t-\mu_{T|y})^\top S^{-1} (dA) \mu_{X|y} \nonumber \\
    &+ \frac{1}{2} \mathrm{tr}( (S^{-1}(t-\mu_{T|y})(t-\mu_{T|y})^\top S^{-1} \nonumber \\
    &\phantom{{}+\frac{1}{2}\operatorname{tr}(} - S^{-1})(dA)\Sigma_{X|y}A^\top ) \nonumber \\
    &+ \frac{1}{2} \mathrm{tr}( (S^{-1}(t-\mu_{T|y})(t-\mu_{T|y})^\top S^{-1} \nonumber \\
    &\phantom{{}+\frac{1}{2}\mathrm{tr}(} - S^{-1})A\Sigma_{X|y}(dA)^\top ).
\end{align}

By definition, for a scalar function $f(A)$ where $A \in \mathbb{R}^{m \times n}$, its total derivative can be written as $df = \sum_{i,j} \frac{\partial f}{\partial A_{i,j}} dA_{i,j}$,
which can be expressed as the Frobenius inner product as follows:
\begin{equation}
    df = \langle \nabla_Af \; , \; dA \rangle_F = \mathrm{tr}((\nabla_Af)^\top dA).
    \label{eq:Frobenius}
\end{equation}

For simplicity, let us define $Q = S^{-1}(t-\mu_{T|y})(t-\mu_{T|y})^\top S^{-1} - S^{-1}$. Using the trace's cyclic property, the result $\mathrm{tr}(A)=\mathrm{tr}(A^\top)$, and the symmetry of the covariance matrix $\Sigma_{X|y}$, we can find both terms
\begin{align}
    \mathrm{tr}(Q(dA)\Sigma_{X|y}A^\top) &= \mathrm{tr}(\Sigma_{X|y}A^\top Q dA) \nonumber \\
    \mathrm{tr}(QA\Sigma_{X|y}(dA)^\top) &= \mathrm{tr}(\Sigma_{X|y}A^\top Q^\top dA).
\end{align}

Additionally, using the fact that matrix $Q$ is also symmetric, we can express $d\log p$ as
\begin{align}
    d \log p(t|y) &= (t-\mu_{T|y})^\top S^{-1}(dA)\mu_{X|y} \nonumber \\
   &+\frac{1}{2}\mathrm{tr}(\Sigma_{X|y}A^\top Q dA) \nonumber \\
   &+\frac{1}{2}\mathrm{tr}(\Sigma_{X|y}A^\top Q^T dA).
   \label{eq:dlop}
\end{align}

Using again the trace's cyclic property, and the identity $a^\top b = \mathrm{tr}(ba^\top)$ for $a,b$ real column vectors, we can write the equation \eqref{eq:dlop} as
\begin{align}
   d \log p(t|y) &= \mathrm{tr}((S^{-1}(t-\mu_{T|y})\mu_{X|y}^\top)^\top dA) \nonumber \\
   &+ \mathrm{tr}((QA\Sigma_{X|y})^\top dA).
\end{align}

Applying again the trace linearity, using the transpose property on addition, and writing in the Frobenius inner product form as equation \eqref{eq:Frobenius}, we finally find
\begin{align}
    d \log p(t|y) &= \mathrm{tr}((S^{-1}(t-\mu_{T|y})\mu_{X|y}^\top + QA\Sigma_{X|y})^\top dA),
\end{align}

which produces
\begin{align}
    \nabla_A \log p(t|y) = S^{-1}(t-\mu_{T|y})\mu_{X|y}^\top + QA\Sigma_{X|y}.
\end{align}

\bibliographystyle{IEEEtran}
\bibliography{refs_submitted}

\end{document}